\documentclass[10pt,conference]{IEEEtran}
\IEEEoverridecommandlockouts
\usepackage{cite}
\usepackage{amsmath,amssymb,amsfonts}
\usepackage{algorithmic}
\usepackage{graphicx}
\usepackage{textcomp}
\usepackage{xcolor}
\usepackage{multirow}
\usepackage{array}
\usepackage{subcaption}
\usepackage{booktabs}
\usepackage{soul}
\usepackage{hyperref}
\def\BibTeX{{\rm B\kern-.05em{\sc i\kern-.025em b}\kern-.08em
    T\kern-.1667em\lower.7ex\hbox{E}\kern-.125emX}}
\begin{document}

\title{A Case Study of Onboarding in Software Teams: Tasks and Strategies}

\author{
\IEEEauthorblockN{An Ju}
\IEEEauthorblockA{\textit{University of California, Berkeley} \\
Berkeley, CA, USA\\
an\_ju@berkeley.edu}
\and
\IEEEauthorblockN{Hitesh Sajnani}
\IEEEauthorblockA{\textit{Microsoft} \\
Redmond, WA, USA\\
hitsaj@microsoft.com}
\and
\IEEEauthorblockN{Scot Kelly}
\IEEEauthorblockA{\textit{Microsoft} \\
Redmond, WA, USA\\
scot.kelly@microsoft.com}
\and
\IEEEauthorblockN{Kim Herzig}
\IEEEauthorblockA{\textit{Microsoft} \\
Redmond, WA, USA\\
kim.herzig@microsoft.com}
}

\maketitle
\newcommand{\company}{Microsoft}

\begin{abstract}
    Developers frequently move into new teams or environments across software companies.
    Their onboarding experience is correlated with productivity, job satisfaction, and other short-term and long-term outcomes.
    The majority of the onboarding process comprises engineering tasks such as fixing bugs or implementing small features.
    Nevertheless, we do not have a systematic view of how tasks influence onboarding.
    In this paper, we present a case study of Microsoft,
    where we interviewed 32 developers moving into a new team and 15 engineering managers onboarding a new developer into their team
    -- to understand and characterize developers' onboarding experience and expectations in relation to the tasks performed by them while onboarding.
    We present how tasks interact with new developers through three representative themes: learning, confidence building, and socialization.
    We also discuss three onboarding strategies as inferred from the interviews that managers commonly use unknowingly, and discuss their pros and cons and offer situational recommendations.
    Furthermore, we triangulate our interview findings with a developer survey ($N=189$) and a manager survey ($N=37$)
    and find that survey results suggest that our findings are representative and our recommendations are actionable.
    Practitioners could use our findings to improve their onboarding processes,
    while researchers could find new research directions from this study to advance the understanding of developer onboarding.
    Our research instruments and anonymous data are available at \url{https://zenodo.org/record/4455937#.YCOQCs_0lFd}.
\end{abstract}


\section{Introduction}
\label{sec:introduction}
Onboarding is a process where new members adjust to their new surroundings and acquire the behaviors, attitudes, and skills necessary to fulfill their new roles and function effectively as a member of a team~\cite{saks2007socialization}.
A large number of developers change teams or start new positions every year but are not productive for months because onboarding can be overwhelming~\cite{williams2003mellon,rollag2005getting}.
A developer often undergoes a high amount of cognitive overload of working on a new system, faces social challenges of belonging to a new team and company, and deals with the pressure of proving oneself in a new environment.
Such stress influences a developer's happiness, retention, and productivity, but could be managed by good onboarding strategies and automated support;
meanwhile, ill-designed onboarding processes could cause a vast waste of company resources and talents.
Therefore, improving developer onboarding crucial~\cite{ko2018msr}.

Developers onboard a new team with tasks, such as software development and maintenance.
Through tasks, new developers learn about the new system, socialize with the team, build confidence, and gain the team's recognition.
However, we lack an understanding of how tasks influence onboarding in \emph{software development teams}.
Hence, we propose the following research question: 
\textbf{RQ1. How do tasks influence a new developer's onboarding experience in software development teams?}

We answer this research question with an exploratory case study, which allows us to examine tasks and onboarding without any existing theory.
Besides, there are two observations that justify this method.
First, onboarding is multidimensional.
There are conventional measurements, such as productivity~\cite{rastogi2017ramp} and socialization~\cite{rollag2005getting}.
However, mental states, such as a developer's confidence, motivation, and stress level~\cite{ellis2015navigating}, are important indicators as well.
Therefore, we view the choice of onboarding tasks as a design decision that balances various consequences and objectives in a specific situation.
Second, onboarding is a long process with many stakeholders.
Onboarding could take over six months~\cite{williams2003mellon,rollag2005getting}, even for developers from within the company~\cite{dagenais2010moving}.
And situations are fast changing in this process.  
For example, the task of documenting a method call may help a new developer to understand the team's workflow and tools;
as the developer learns, such simple tasks quickly become inappropriate, and the developer needs more challenging tasks to keep motivated. 
Therefore, we believe the onboarding process is continuously evolving.

We also aim to uncover some commonly used \emph{strategies} in practice to assign tasks when onboarding new developers.
Our understanding of tasks and onboarding will help us analyze these strategies, and we can further provide practical suggestions for researchers and practitioners to understand and improve onboarding processes.
Formally, we propose the following research question:
\textbf{RQ2. What task-assignment strategies should be used in order to facilitate a positive onboarding experience in software engineering teams?}

We use interviews and surveys to answer the two research questions,
where we get exploratory findings from interviews and test these findings with surveys.
We interviewed 32 developers who had recently joined a new team and 15 managers who had recently had a new developer,
allowing us to collect inputs from both sides of onboarding processes.
Our interviews revealed how software development tasks influence some important aspects of onboarding processes, such as learning, socialization, and the new developer's confidence.
Besides, we uncovered three onboarding strategies that managers commonly use to onboarding new developers.
We use surveys to get feedback from a broader audience regarding interview findings.
We collected survey responses from 189 developers and 37 managers.
Survey results suggest that our interview findings are representative of onboarding experiences and that the three strategies can represent a majority of onboarding strategies in practice.

Our case study leads to several recommendations that managers or organizations may use to improve onboarding processes.
These recommendations are validated by our surveys as actionable and useful.
Managers can also benefit from our discussions about the three onboarding strategies.
We compare the pros and cons of the three strategies so that practitioners can choose their strategy that best fits their scenarios.

This paper is organized in the following way.
Section~\ref{sec:method} describes our method.
Section~\ref{sec:results} presents our interview findings.
Section~\ref{sec:survey} shows survey results that validate our interview findings in Section~\ref{sec:results}.
Section~\ref{sec:related} discusses our work's connections with previous studies.
Section~\ref{sec:threats} addresses some threats of validity,
and Section~\ref{sec:conclusion} concludes this paper with some future research directions.

\begin{figure*}[ht]
    \begin{center}
    \includegraphics[width=\linewidth]{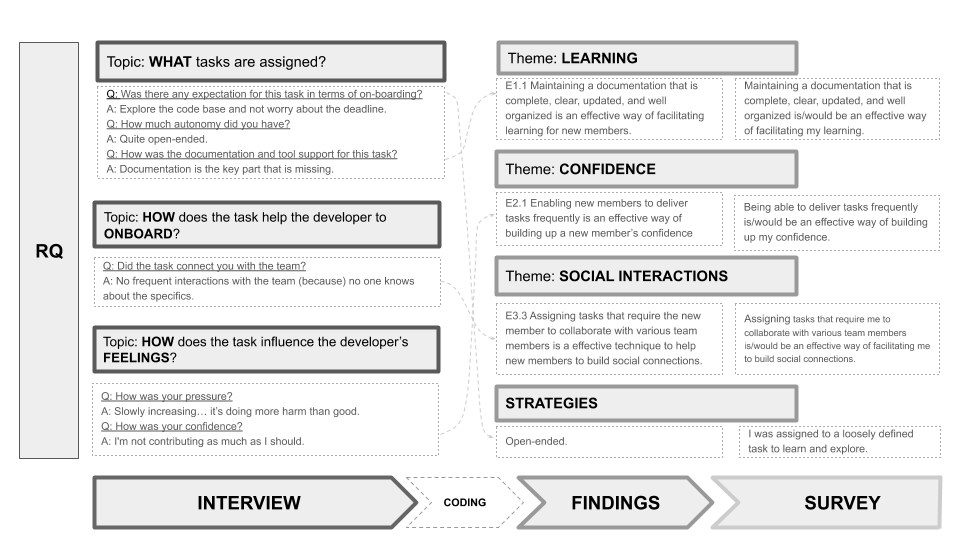}
    \end{center}
    \caption{An illustrative overview of our research process. We get exploratory findigns from interviews and test these findings with surveys. Interview questions are designed to measure tasks and each task's role in onboarding. Interview transcripts are coded into themes (findings); Table~\ref{tab:findings_recommendations} gives a complete view of our interview findings and Section~\ref{sec:coding} has more details about the coding process. Surveys triangulate our findings with a broader audience. Survey questions are derived directly from our interview findings.}
    \label{fig:research_process}
\end{figure*}

\section{Methods}
\label{sec:method}
Fig.~\ref{fig:research_process} is an overview of our two-step research method.
Supplementary materia (available at \url{https://zenodo.org/record/4455937#.YCOQCs_0lFd}) has interview guides and survey questions.

\subsection{Subjects and Research Site}
\label{sec:subjects_and_research_site}

\begin{table*}[ht]
    \caption{Interview subjects.  Transfer means new developers from another team within \company{}. Division A and Division B are two divisions under \company{} that have different businesses, cultures, and onboarding practices.}
    \begin{center}
    \begin{tabular}{m{0.1\linewidth}m{0.1\linewidth}|m{0.1\linewidth}m{0.1\linewidth}|m{0.1\linewidth}m{0.1\linewidth}|m{0.1\linewidth}m{0.1\linewidth}}
          & & \multicolumn{2}{c|}{Gender} & \multicolumn{2}{c|}{Division} & \multicolumn{2}{c}{Background} \\
                        & Sum   & Male          & Female    & Division A         & Division B         & New Hire      & Transfer \\\toprule
         Developers     & 32    & 23 (72\%)     & 9 (28\%)  & 22 (69\%)     & 10 (31\%)     & 21 (66\%)     & 11 (34\%) \\
         Managers       & 15    & 13 (87\%)     & 2 (13\%)  & 12 (80\%)     & 3 (20\%)      & -             & -         \\\bottomrule
    \end{tabular}
    \end{center}
    \label{tab:interview_subjects}
\end{table*}

\begin{table*}[ht]
    \caption{We sent surveys to these developers and managers and received 189 replies for the developer survey and 37 replies for the manager survey. Seniority is inferred from employee titles.}
    \begin{center}
    \begin{tabular}{m{0.1\linewidth}m{0.1\linewidth}|m{0.1\linewidth}m{0.1\linewidth}|m{0.1\linewidth}m{0.1\linewidth}|m{0.1\linewidth}m{0.1\linewidth}}
          & & \multicolumn{2}{c|}{Seniority} & \multicolumn{2}{c|}{Division} & \multicolumn{2}{c}{Background} \\
                        & Sum     & Junior          & Senior    & Division A         & Division B         & New Hire      & Transfer \\\toprule
         Developers     & 1629    & 978 (60\%)     & 651 (40\%)  & 1048 (64\%)     & 581 (36\%)     & 1097 (67\%)     & 532 (33\%) \\
         Managers       & 754    & 386 (51\%)     & 368 (49\%)  & 507 (67\%)     & 247 (33\%)      & -             & -         \\\bottomrule
    \end{tabular}
    \end{center}
    \label{tab:survey_subjects}
\end{table*}

This study was conducted at the US division of \company{}, a major tech company with more than 100,000 employees.
\company{} has thousands of new hires and internal transfers every month making it a good research site to study onboarding.
Besides, \company{} has multiple divisions, each with relatively distinct businesses, cultures, and onboarding practices.
We conducted our study from two divisions (Division A and Division B) to improve the generalizability of our findings.
Division A focuses on service and software development, while Division B has both software and hardware businesses.
Both divisions have a large number of software engineering teams.

We focus on software development teams;
a team has one engineering manager (manager) and several software developers (developers) with a goal of delivering products and services.
All teams are atomic; there is no sub-team within a team.
A manager's goal is two-fold:
(i) help team members grow in their roles;
(ii) provide technical leadership to the team, keeping the developers on track to deliver priorities and maintaining good team health and productivity.
Accordingly, a manager's job includes planning and negotiating, assigning tasks, evaluating, and resolving inter and intra-team conflicts.
Managers decide how tasks are assigned to developers within a team and \company{} has no standard strategy for task assignment.
For example, some managers choose to follow Agile methods and developers get their tasks following the order of priority,
but some managers may choose to assign tasks directly to developers.
In our interviews, all developers clearly understand the meaning of managers.

Our analysis is grounded on tasks.
A typical task is an item from the team's task board, defined and scoped to 1-4 weeks;
some exploratory tasks are time-boxed to 1-2 weeks.
In our interviews, some developers mentioned tasks customized for onboarding, such as fixing and updating documentation;
these tasks are also defined and time-boxed.
We found that both developers and managers understand the meaning of tasks well.

Company \company{} maintains a \emph{Person} database that contains historical details about each employee's personal (e.g., date of joining, title, job description, seniority level and years in the company), the team (e.g., product group, team members, manager, and leadership chain), and geographic (city, building) details.
To find subjects, researchers queried the \emph{Person} database to identify: developers who had recently
joined a new team
, and developer managers who had had a new developer joined recently.
For interviews, we sampled developers who joined a new team between April 1st, 2019, to July 1st, 2019 (in the past three months of this study) and developer managers
who had a new developer between January 1st, 2019 to July 1st, 2019 (in the past six months of this study). 
For surveys, we set 6 months as the threshold for both developers and developer managers.
We filtered results to keep only software developers or software developer managers who work at the same location (US Headquarter) as researchers.
This process yielded 397 developers and 1167 developer managers for interviews and 1629 developers and 754 developer managers for surveys.
Some demographics of interview and survey subjects are presented in Table~\ref{tab:interview_subjects} and Table~\ref{tab:survey_subjects}.

\subsection{Interviews}
We conducted a pilot study comprising of six developer interviews prior to the formal study.
Pilot interviews were unstructured, focusing on the developer's onboarding experience and tasks.
From the pilot study, we found constructs~\footnote{We use constructs to refer to artifacts, such as documentations, processes, such as meetings, mental feelings, such as confidence, and metrics, as learning.} that were closely related to onboarding; 
these findings are used to design interview guides.

\subsubsection{Design}
To understand how tasks influence onboarding, we focus on 1) the context of onboarding, such as the developer's prior experience, 2) tasks, such as a task's complexity and expected results, and 3) constructs influenced by tasks, such as learning and confidence.
For developers, we ask participants to describe a task and ask follow-up questions about how the task influence the other constructs.
For managers, we ask participants to describe how they assign tasks, such as their expectations and evaluations.
Interview questions can be found in the \href{https://zenodo.org/record/4455937#.YCOQCs_0lFd}{supplementary material}.

\subsubsection{Administration}
We randomly sampled subjects from the pool described in Section~\ref{sec:subjects_and_research_site} and contacted each subject through emails to schedule the interview.
Each interview took 30-60 minutes, administrated by at least two researchers and audio/video recorded.
One researcher read informed consent to the participant and collected the participant's verbal consent before the interview. 
Interviews were led mostly by one researcher and transcribed by the other.
Researchers consolidated interview notes right after the interview.

\subsubsection{Analysis}
\label{sec:coding}
\begin{figure*}
    \begin{center}
    \includegraphics[width=\linewidth]{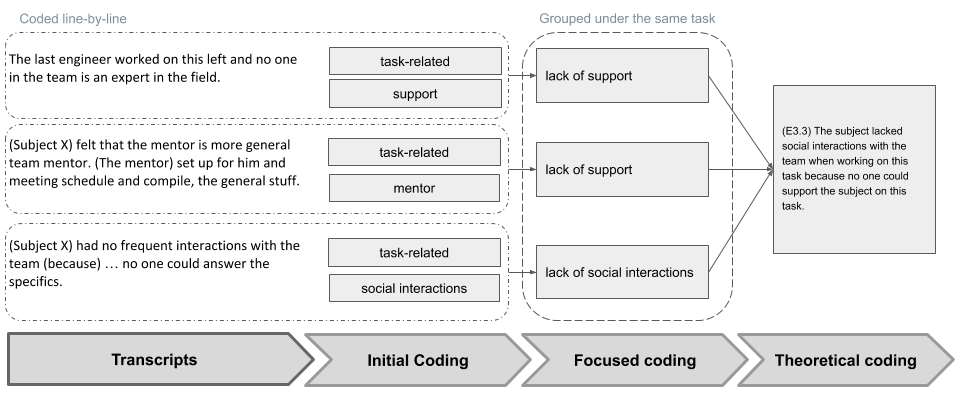}
    \end{center}
    \caption{An illustrative overview of the coding process.}
    \label{fig:coding_process}
\end{figure*}
We followed the coding process from constructivist grounded theory~\cite{charmaz2014constructing,stol2016grounded} to code interviews.
One researcher finished the coding process through a spreadsheet of transcript lines.
Researchers consolidated findings after each coding step and decided focuses for the next step together.
After focused coding, researchers discussed the emerged themes to be used for theoretical coding with other domain experts, including other researchers, developer team leaders, and senior managers from \company{} that specifically work in the domain of improving developer productivity. While this exercise did not result any significant changes in the emerged themes but helped induce more confidence in the themes and avoid bias.

\textbf{Initial coding.}
We coded interview transcripts line-by-line into topics, tasks, constructs.
We used three topics (\emph{pre-task} for contexts of tasks, \emph{task-related} for specific tasks, \emph{task sequences} for the ordering of tasks)for developer interviews
and two additional topics (\emph{goals} and \emph{measurements}) for developer interviews.
Furthermore, we assigned each \emph{task-related} line a descriptor, such as \emph{priority} and \emph{complexity}.
An initial list of descriptors was obtained from previous reports, expert suggestions, and our pilot interviews, and continuously revised in the coding process.

\textbf{Focused coding}
We coded at task level in focused coding, characterizing each task with a reduced set of codes, such as the quality of support and social interactions, as presented in Fig.~\ref{fig:coding_process}.
The list of codes used in focused coding was identified as the most relevant and frequent by researchers in initial coding;
a complete list of codes and their meanings could be found in the \href{https://zenodo.org/record/4455937#.YCOQCs_0lFd}{supplementary material}.


\textbf{Theoretical coding}
We focus on three themes in theoretical coding: learning, confidence, and social interactions.
The three themes frequently appear in interviews and are representative of onboarding experiences.
Therefore, presenting tasks and their interactions with onboarding under the three themes covers artifacts (learning), mental status (confidence), and interpersonal dynamics (social interactions).
We believe other interactions, such as how tasks motivate new members, are similar to one of the three themes.
Besides the three themes, we analyzed each developer's onboarding experience and each manager's onboarding design to understand onboarding strategies.
We have summarized three onboarding strategies from analyzing tasks coded in focused coding.
We further analyze the pros and cons of each onboarding strategy in theoretical coding.

\subsection{Surveys}
\subsubsection{Design}
\textbf{After} we coded interviews, we created surveys to collect broad feedback to some specific interview findings and onboarding practice recommendations, and the three onboarding strategies.
As Fig.~\ref{fig:research_process} suggests, the developer survey contains questions derived from our interview findings in Table~\ref{tab:findings_recommendations} and Section~\ref{sec:strategies}.
It has two sections.
1) The first section asks developers to choose a strategy, from the three strategies listed in Section~\ref{sec:strategies}, that best describes their recent onboarding experience or ``None of the above'';
developers can explain their choice with an open-response questions.
2) The second section has four question groups, corresponding to three themes and a theme of \emph{mentors};
we add \emph{mentors} because mentors are critical for learning, building up confidence, and creating social connections.
Each question group contains Likert-scale questions that ask developers to rate their level of agreement to some statements,
ending with two open-response questions for developers to explain their choices and provide recommendations.
For example, participants are asked
\begin{quote}
    Based on you recent on-boarding experience, to what degree do you agree or disagree with the following statements about learning?
    
    Maintaining a documentation that is complete, clear, updated, and well organized is/would be an effective way of facilitating my learning.
\end{quote}
which validates the interview finding E1.1 in Table~\ref{tab:findings_recommendations} directly.
In addition to the two sections, we collect some background information prior to the first section.
The developer survey used in this study can be found in the \href{https://zenodo.org/record/4455937#.YCOQCs_0lFd}{supplementary material}.

The manager survey validates our practice recommendations and the three strategies.
It contains two sections.
1) In the first section, managers are asked to choose one strategy that best describes how they onboard a junior or senior developer, respectively;
they are presented with the three strategies as well as a choice of "none of the above" and are encouraged to explain their choice with an open-ended question.
2) The second section asks managers to rate how likely they would use recommendations from Table~\ref{tab:findings_recommendations} to improve onboarding;
they may choose ``Already in practice'' if the recommended practice is being used.
For example, participants are asked
\begin{quote}
    Do you believe that the following practices that could facilitate learning would improve a new member's on-boarding experience in your team?
    
    Create and/or maintain a new member-friendly documentation.
\end{quote}
which validates the recommendation A1.1 in Table~\ref{tab:findings_recommendations} directly.
In addition to the two sections, we collect some background information prior to the fist section.
The manager survey used in this study can be found in the \href{https://zenodo.org/record/4455937#.YCOQCs_0lFd}{supplementary material}.

We take the following steps to ensure our survey is valid and not biased.
First, we conduct two rounds of in-person pilot studies with 5 recently onboarded developers and managers who recently onboarded a new developer in their team and used their feedback to remove ambiguity and biases in our initial design. The pilot studies' participants do not participate in the actual survey. Second, to ensure the participant's understanding, we define important terms in the context of our survey before the questions.
All questions and definitions are included in the \href{https://zenodo.org/record/4455937#.YCOQCs_0lFd}{supplementary material}.

\subsubsection{Administration}
We sent surveys to 1629 developers and 754 developer managers identified from the pool, as described in Section~\ref{sec:subjects_and_research_site} via emails;
interview subjects were excluded from surveys.
We received 189 replies for the developer survey, and 37 replies for the manager survey.


\section{Interview Findings}
\label{sec:results}


This section presents the interview findings.
To answer RQ1, our interviews have revealed fifteen constructs that are relevant to onboarding, including context factors such as the developer's prior experience,
onboarding metrics such as learning and social interactions, mental status such as confidence and motivation, and processes such as stand-up meetings.
Three constructs are particularly critical to onboarding experience:
learning, which measures how the new developer gains knowledge about the new team~\cite{dagenais2010moving};
confidence, which is closely related to how the new developer \textbf{feels} about the onboarding process;
and social interactions, which is a critical indicator that managers use to measure onboarding~\cite{saks2007socialization}.
Section~\ref{sec:expectations_and_task} will elaborate on how learning, confidence, and socialization are influenced by tasks,
and our key findings are presented in Table~\ref{tab:findings_recommendations}.
Furthermore, to answer RQ2, Section~\ref{sec:strategies} shows three onboarding strategies that have emerged from interviews.
Across this section, we refer to interviewed developers with P and interviewed developer managers with M.


\subsection{Learning, Confidence, and Social Connections}
\label{sec:expectations_and_task}


\begin{table*}[ht]
    \caption{Interview findings and recommendations derived from the findings.
    The numbers for Experiences are calculated from the developer survey as the percentage of developers who have answered ``Agree'' or ``Strongly agree''.
    The numbers for Practices are calculated from the manager survey as the percent of managers who have answered ``Very likely'', ``Somewhat likely'', or ``Already in practice''.
    Creating a psychologically safe environment~\cite{edmondson1999psychological} is a broad topic that is beyond the scope of this paper.
    Readers can refer to other studies such as Google's Project Aristotle~\cite{duhigg2016google}.
    }
    \begin{center}
    \begin{tabular}{m{0.04\linewidth}|m{0.48\linewidth}m{0.03\linewidth}|m{0.32\linewidth}m{0.03\linewidth}}
         Topic &  Experiences &  & Practices & \\\toprule
         \multirow{4}{\hsize}{\rotatebox[origin=c]{90}{Learning}} & 
         \textbf{E1.1}: Maintaining a documentation that is complete, clear, updated, and well organized is an effective way of facilitating learning for new members. & 90\% & \textbf{A1.1}: Create and/or maintain a new member-friendly documentation. & 93\% \\\cmidrule{2-5}
         & \textbf{E1.2}: Creating a safe and supportive environment where new members can ask questions freely is an effective way of facilitating learning for new members. & 98\% 
         & Create a psychologically safe environment. & \\\cmidrule{2-5}
         & \textbf{E1.3}: Team meetings, such as daily stand-ups and sprint planning meetings, are useful opportunities for new members to learn the team's big picture. & 78\%
         & \textbf{A1.3}: Include new members in the relevant business and technical meetings. & 76\% \\\cmidrule{2-5}
         & \textbf{E1.4}: Explaining the team's big picture (such as visions, missions, and plan) is an effective way of facilitating learning for new members. & 92\%
         & \textbf{A1.4}: Run onboarding sessions in the first few days to explain the big picture to the new member. & 89\% \\\midrule
         \multirow{5}{\hsize}{\rotatebox[origin=c]{90}{Confidence}} & \textbf{E2.1}: Enabling new members to deliver tasks frequently is an effective way of building up a new member's confidence. & 88\%
         & \textbf{A2.1}: Use a sequence of quick (proper to the new member's seniority) tasks for new members to on-board. & 84\% \\\cmidrule{2-5}
         & \textbf{E2.2}: Providing frequent positive confirmation and trust from the team is an effective way of building up a new member's confidence. & 91\%
         & \textbf{A2.2}: Set frequent checkpoints to give continuous positive confirmation to the new member. & 92\% \\\cmidrule{2-5}
         & \textbf{E2.3}: New members build up their confidence as they learn more about the team's landscape. & 92\%
         & \textbf{A2.3}: Set up dedicated learning time/sessions for new members. & 81\% \\\cmidrule{2-5}
         & \textbf{E2.4}: Explaining a task's big picture and clarifying the specifications is an effective way of building up confidence for new members. & 86\%
         & \textbf{A2.4}: Make sure the new member clearly understand the specifications of the task. & 92\% \\\cmidrule{2-5}
         & \textbf{E2.5}: Creating an environment where new members feel safe and supported is an effective way of building up confidence for new members. & 95\%
         & Create a psychologically safe environment. & \\\midrule
         \multirow{3}{\hsize}{\rotatebox[origin=c]{90}{Socialization}} & \textbf{E3.1}: New members build up social connections with the team faster and easier when they can interact frequently with the manager and mentor. & 80\%
         & \textbf{A3.1}: Set up frequent one-on-one meetings with a new member in the first few weeks. & 95\% \\\cmidrule{2-5}
         & \textbf{E3.2}: Having a low-cost channel for new members to seek help from the team is an effective way of facilitating a new member to build up social connections with the team. & 87\%
         & \textbf{A3.2}: Use social events to create a safe environment for the new member. & 92\% \\\cmidrule{2-5}
         & \textbf{E3.3}: Assigning tasks that require the new member to collaborate with various team members is a effective technique to help new members to build social connections. & 89\%
         & \textbf{A3.3}: Assign tasks that require the new member to work with different people from the team. & 86\% \\\bottomrule
    \end{tabular}
    \end{center}
    \label{tab:findings_recommendations}
\end{table*}

Ten managers considered independence as a \emph{final checkpoint} for onboarding.
Independence is indicated by answering customer questions, driving complex features, contributing new ideas, and more.
We focus on three aspects: learning, confidence, and social connection,
because they are necessary for independence and are frequently mentioned in manager interviews.
Table~\ref{tab:findings_recommendations} summarizes our findings and our recommendations for practitioners to facilitate the new member's learning, confidence, and social connections.

We omit most direct quotes in this section due to the page limit.
Readers can find more quotes and details in the \href{https://zenodo.org/record/4455937#.YCOQCs_0lFd}{supplementary material}.

\subsubsection{Learning}
New developers need to learn everything about the team: code repositories, business values and customers, development tools and processes, and responsibilities of each individual member and some other close teams.
Developers frequently learn from their tasks.
Out of 61 \emph{task items}, 41 (67\%) led to learning.
As P14 put it: ``there is so much to learn, so any task assigned to me would have created a learning experience''.
Task-related documentations and team support are major knowledge channels for new members.
Besides, team meetings also provide learning opportunities.

\textbf{Task-related documentation.} 
Task-related documentations, including documents, presentations, tutorials, and demos, are primary sources of information.
Twelve developers mentioned that they learned from documentations.
Developers seek answers from documentations before asking for the team's help and they feel frustrated when documentation cannot provide answers.
Some developers mentioned the importance of indexing documents in particular, suggesting that making information organized and easy-to-find can improve onboarding experiences.
Therefore, the team should assign new developers tasks that are supported by clear, complete, updated, and well-organized documentations, so that new members can quickly locate relevant information and avoid inaccurate or incomplete information.

\textbf{Team support.}
Team support is critical for learning.
Not all information is available as documentations, so new developers have to seek the team for help.
In particular, mentorship is a typical and effective form of team support.
Mentors help new developers solve problems, master development tools and processes, and find useful resources;
they also work as a bridge so that new developers can seek help from others in the team.
Compared to other team members, new developers find it easier to seek help from a dedicated person, like mentors.
Therefore, assigning task-support mentors is an efficient method to promote learning.

\textbf{Technical meetings.}
Developers benefit from attending technical meetings, although these meetings are not designed for learning.
These meetings provide new information about the team's work, its connections to other teams, its workflow, and its division of responsibilities.
Some meetings are conversational, where new developers get their questions answered.
For example, standups are opportunities for new developers to ask task-related questions.
Team meetings are particularly useful when teams have many new developers.
In those cases, participants spend more time on learning and understanding,
while in some other cases, new developers may find information from these meetings overwhelming.

\textbf{Learn the task's big picture first.}
Learning a task's contexts facilitate a new developer's learning.
14 developers found it helpful that their managers presented big pictures before assigning tasks,
and some developers struggled because they did not see the big picture first.
For example, P21 had problems seeking help from the team not knowing the big picture.
Focusing on the big picture and avoiding details prevents information overload because new developers can filter unnecessary information.
Therefore, we recommend managers to explain the task's context and its connections to the rest of the team before assigning a task.

\subsubsection{Confidence}
New developers become more confident in their team roles as they onboard.
Confidence is necessary for independence, but low confidence is common for new members. 
Five developers mentioned that they had low confidence due to unfamiliar technologies and processes.
Four sources of confidence have emerged from interviews: finishing a task, confirmation and trust, learning, and team support.

\textbf{Finishing a task.}
Finishing a task boosts confidence.
New developers prove that they can deliver values for the team and they get more confidence about the team's development process.
On the other hand, having no progress hurt confidence.
Slow progress causes new developers to feel pressure and low confidence, even when new developers know that the team does not expect them to finish the task soon.
Therefore, assigning tasks that could be quickly solved could boost confidence of new developers.

\textbf{Receiving positive feedback on performance.}
New developers gain confidence when they feel the team's confirmation and trust.
There are many indicators of trust and confirmation:
some developers feel confident when other team members start to ask him/her for help;
others feel confirmation when they are nominated to handle customer requests.
Meanwhile, negative judgments about a new developer's abilities could hurt confidence.
P17 felt less confident when her co-worker made jokes about her being too slow even when she knew the co-worked did not mean it.
Getting early positive feedback usually leads to good onboarding experiences;
for example, senior developers can receive confirmation and trust from the team faster than junior developers because they have advantages in the team on their tasks, and senior developers also have better onboarding experiences than junior developers in our interviews.

\textbf{Clear learning outcomes.}
New developers become more confident as they become more familiar about the team and their work.
P19 and P20 mentioned that systematic learning on task-related topics boosted their confidence.
Clarifying task specifications is also a form of learning that improves confidence.
P8 said that ``understanding the requirements clearly is related to gaining confidence'' because his task felt more normal knowing the requirements.
Therefore, we recommend managers to emphasize on what the new developer could learn from assigned tasks and set clear expectations of learning gains when assigning tasks.

\textbf{Peer support.}
New developers are confident when they have the team support working on a task.
As P25 put it: ``I'm aware that there are plenty of things that I don't know, but I know there are others who know \dots and I'm confident that there are people where I can get help''.
Support may come from the manager and mentor specifically.
New developers have more confident when they know there is a specific person from whom they can seek help.
If new developers are paired with a senior developer to work on a task, they typically have a very strong confidence.
Therefore, assigning tasks that are supported by the team and making the support accessible to new developers could improve onboarding.

\subsubsection{Socialisation}
Successful onboarding means positioning oneself in the team and owning a part of the team's work.
We have found three types of social interactions: interactions with mentors, interactions with teammates, and interactions due to collaboration and dependency of tasks.

\textbf{Mentors.}
Mentors are the first person that new developers contact when they have questions.
Out of 18 participants who had a mentor or ``onboarding buddy'', 14 found mentors helpful.
Besides, people who did not have a mentor believed that having a mentor would improve their onboarding experiences.
For example, P9 reported that ``I wish I had someone that I could have sat together with'';
Mentors have a wide range of responsibilities from setting up environments and introduce the team, to answering questions and assigning tasks.
One important responsibility is to reduce barriers for help-seeking.
Mentors are accessible, and new members can ask mentors ``dumb'' questions freely.
Accordingly, some managers mentioned that communications skills are important for mentors.

\textbf{Seeking help from teammates.}
New developers frequently seek help from teammates while working on their tasks.
Through help-seeking, new developers build social connections with the team.
Out of 38 \emph{task items} where social interactions were discussed, 29 led to a strengthened connection between the developer and the team.
However, new developers might feel hesitant to ask questions because they feel unsafe, or they do not want to bother others.
As Rollag et al. pointed out, managers should not expect new members to fend for themselves~\cite{rollag2005getting}.
If help-seeking is minimal, social interactions also decrease.
Therefore, having a low-cost or dedicated channel for help-seeking could promote social interactions.
For example, P4 had a good onboarding experience because he could ask questions in a dedicated online channel and got help from experienced developers fast.
This dedicated channel was a low-cost approach for P4 to seek help so that he did not have to figure out whom to ask, and there were no social barriers for him to ask related questions.

\textbf{Collaboration and dependency of tasks.}
Tasks that involve interactions and collaborations with multiple stakeholders force new developers to establish new social connections.
Some tasks require collaborations with another team member, but some tasks require cross-team and cross-domain collaborations.
New developers who are sensitive to other people's working styles generally feel happy to work with others.
However, some new developers may prefer tasks that are less socially complex.
In particular, waiting is common when collaborating with others and waiting has a negative impact on the motivation and confidence of new developers.
For example, P23 worked with other teams on a new feature, but could not get timely feedback because the other teams would not review pull requests.
In general, tasks that require cross-team or cross-domain collaborations are challenging for new developers and should be used with caution. 

\subsubsection{Answering RQ1.}
Our interviews suggest that onboarding is a long-term process shaped by tasks and that a task's influence on onboarding is compounded with the its contexts.
For example, tasks with high uncertainties could facilitate learning and social interactions with team support, but could hinder onboarding when no support is available.
We present how tasks influence learning, confidence, and socialization and we provide some recommendations for practitioners to improve their onboarding processes.
Our findings and recommendations are presented in Table~\ref{tab:findings_recommendations}.


\subsection{Onboarding Strategies}
\label{sec:strategies}
This section presents three onboarding strategies that we have summarized from interviews:
\begin{itemize}
    \item \textbf{Simple-Complex} is an approach where managers gradually increase the task's complexity.
    \item \textbf{Priority-First} is an approach where managers follow the order of priority.
    \item \textbf{Exploration-Based} is an approach where managers assign tasks that are under-defined and uncertain.
\end{itemize}

\subsubsection{Simple-Complex}
Simple-Complex is widely used.
12 managers used this strategy, while 14 developers were onboarded with this strategy.

In Simple-Complex, managers increase the task's complexity over time.
M8 described a typical Simple-Complex:
\begin{enumerate}
    \item First week: use simple tasks (bugs or configuration settings) to get familiar with the development process.
    \item The next 2 or 3 weeks: get oriented in the codebase with small bugs or simple features.
    \item The next 3 to 9 months: become an expert in one domain with more bugs and features.
    \item Eventually: become an expert who can handle the design of a new feature.
\end{enumerate}
This progressive approach is consistent with expectations that managers have when onboarding new members.
As a comparison, M14, who also used Simple-Complex, described his expectations:
\begin{enumerate}
    \item First week: learn tools and source code.
    \item Second week: learn the customer's perspectives.
    \item 3-4 months: they can design something or add a new capability in a thoughtful way.
    \item Six months: they should be able to work independently.
\end{enumerate}
With Simplex-Complex, managers choose tasks for the developer to achieve expected learning goals.
It creates a smooth learning experience for new members.
Besides, as finishing tasks builds up confidence, the developer could maintain high confidence or build up confidence fast.

Tasks in this stage are simple and well-defined.
Some managers would look for high-priority tasks, because ``it gives visibility to the person'' (M11).
But others did not care about priority or wanted to avoid high-priority tasks.

There are two variants based on how the manager assign tasks after 3 to 9 months.
\begin{itemize}
    \item \textbf{Depth-First} lets the developer become expert in one domain and expand knowledge gradually to other domains. 
    \item \textbf{Breadth-First} exposes the developer to multiple parts first before assigning more in-depth tasks. 
\end{itemize}
Depth-First is more common in our interviews.
Owning a domain improves confidence and promotes social interactions, which could justify Depth-First.
On the other hand, Breadth-First emphasizes the developer's interest and motivates developers with the freedom to explore.
It also helps the developer to build a broad social network with the team.
M9 mentioned: ``those tasks usually make it so that people will meet more team members to work with'',
and P7 mentioned that he liked Breadth-First because ``whenever they talk about something, I know some part of it''.
Further studies could measure and compare the two approaches.

\subsubsection{Priority-First}
Two managers assign tasks following the backlog's order, which we call Priority-First.
M1 described the approach as ``all features are broken down into tasks, and then all people pick up tasks organically''; M2 explained that ``no tasks/projects are new guy projects''.
In both cases, developers get the task with the highest priority.
Twelve developers were onboarded with this strategy.

Priority-First allows the developer to generate values immediately for the team and thus is commonly used by teams under pressure.
Out of six developers who were in a team under pressure, four were onboarded with Priority-First.
In cases such as P14, the developer worked closely with the team on urgent tasks;
although P14 had to learn piece-by-piece as he worked through tasks, he felt positive about his onboarding experience.
For P14, the team's high-pressure and highly focused situation created a safe and energetic learning space.
He reported that he was highly motivated because his coworkers were highly focused;
besides, he felt safe to make mistakes and asked for help because the team was ``doing things too quickly to worry about petty things like people being insulted or worrying about how they took something''.

Although P14 mentioned some benefits, eight developers onboarded with Priority-First mentioned that their tasks were challenging.
Challenging tasks early in the onboarding process could spur fears and low confidence.
P30 said that she had zero confidence for an urgent task, and she was wondering why her manager did that.
Meanwhile, good team support can counter some negative influences.
M1's team practiced pair programming that allowed the developer to interact heavily with an experienced developer;
M2 also paired the developer with an experienced developer on the first few tasks.

\subsubsection{Exploration-Based}
Exploration-Based is a strategy where managers assign loosely defined tasks to the developer.
One manager mentioned that he would assign a task with high uncertainties for senior developers and six developers were onboarded with Exploration-Based.

Managers use Exploration-Based to onboard senior developers.
P24 was aware that some specifications were up to him because he was an experienced engineer.
However, this strategy was also used to onboard junior developers.
For example, P5 was assigned to a technical spike, and P29 was assigned to prototyping a new tool.
Both tasks were outside the team's core product so that new developers could explore without time pressure or worrying about breaking the production code.
Their managers made clear that the task was meant for them to learn.
We found such a strategy might have negative influences on new developers.
At the time we had the interviews, neither P5 and P24 had delivered the assigned task.
P5, in particular, expressed that he felt pressure because he could not deliver as frequently as others.

New teams are more likely to use Exploration-Based to onboard developers.
5 out of 9 developers who were reportedly from a new team were onboarded with Exploration-Based.
New teams have more unspecified tasks, so they are more likely to use this strategy.
Besides, new members feel more comfortable with Exploration-Based in new teams.
P19 mentioned that he felt good because everyone in the team was learning together.
P31 said that people in a new team were aware that new members could be lost so they would provide the necessary help.
Both P19 and P31 were able to build broad social connections with the team through the task.

\subsubsection{Answering RQ2.}
Our interviews have surfaced three onboarding strategies: Simple-Complex, Priority-First, and Exploration-Based.
Simple-Complex is a good strategy for onboarding junior developers;
it gradually prepares new developers, keeping them confident and motivated.
Meanwhile, Exploration-Based works better for experienced developers;
it provides freedom to learn about the team and build an ownership.
Priority-First is used by teams with special organization or special needs;
Agile teams, new teams, and teams with a tight schedule may use this strategy.


\section{Survey Results}
\label{sec:survey}

We sent the developer survey to 1629 developers and collected 189 responses (11.6\%).
We sent the manager survey to 754 manager and collected 37 responses (4.9\%).
Anonymous survey responses are available at \url{https://zenodo.org/record/4455937#.YCOQCs_0lFd}.

\begin{figure*}[ht]
    \centering
    \begin{subfigure}[b]{0.32\linewidth}
        \includegraphics[width=\textwidth]{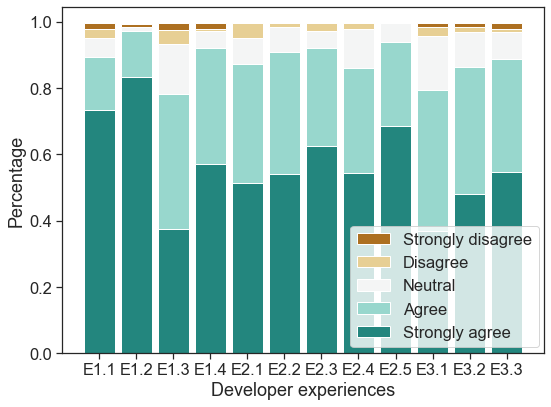}
        \caption{Developer survey results}
        \label{fig:developer}
    \end{subfigure}
    \hfill 
    \begin{subfigure}[b]{0.32\linewidth}
        \includegraphics[width=\textwidth]{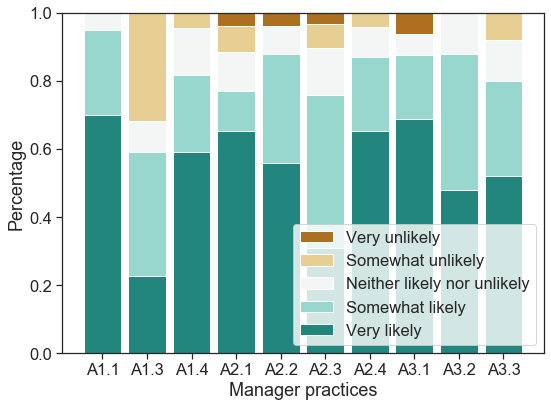}
        \caption{Manager survey results}
        \label{fig:manager}
    \end{subfigure}
    \hfill
    \begin{subfigure}[b]{0.32\linewidth}
       \includegraphics[width=\textwidth]{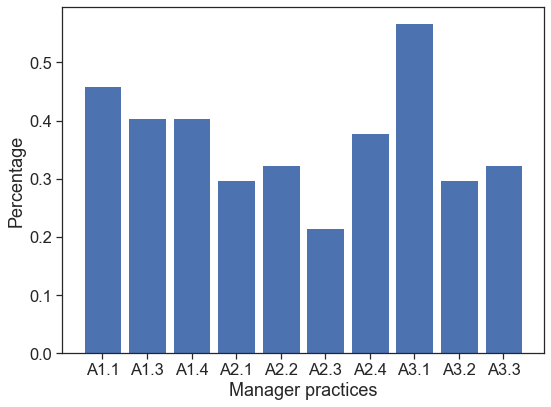} 
       \caption{Practices in use}
       \label{fig:manager_in_practice}
    \end{subfigure}
    \caption{Survey results about developer experiences and recommended practices.  Labels are listed in Table~\ref{tab:findings_recommendations}.  All questions in the developer survey have at least 172 replies ($>91\%$).  All questions in the manager survey have at least 36 relies ($>92\%$).}
    \label{fig:survey_results}
\end{figure*}

\begin{figure*}[ht]
    \centering
    \begin{subfigure}[b]{0.32\linewidth}
        \includegraphics[width=\textwidth]{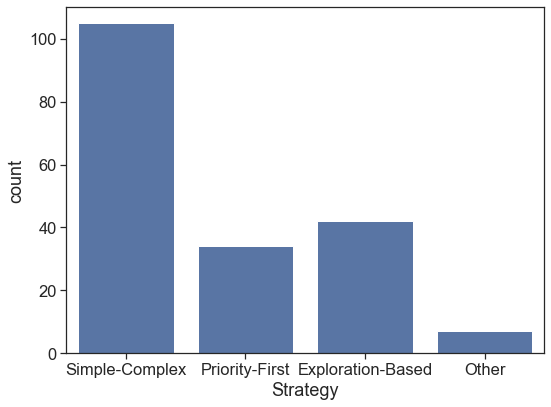}
        \caption{Developer-reported onboarding strategies ($N=188$).}
        \label{fig:developer_strategies}
    \end{subfigure}
    \hfill
    \begin{subfigure}[b]{0.32\linewidth}
        \includegraphics[width=\textwidth]{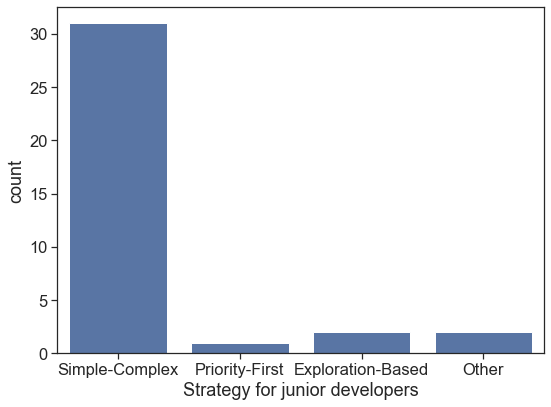}
        \caption{Manager-reported onboarding strategies for junior developers ($N=36$).}
        \label{fig:manager_strategies_junior}
    \end{subfigure}
    \hfill
    \begin{subfigure}[b]{0.32\linewidth}
       \includegraphics[width=\textwidth]{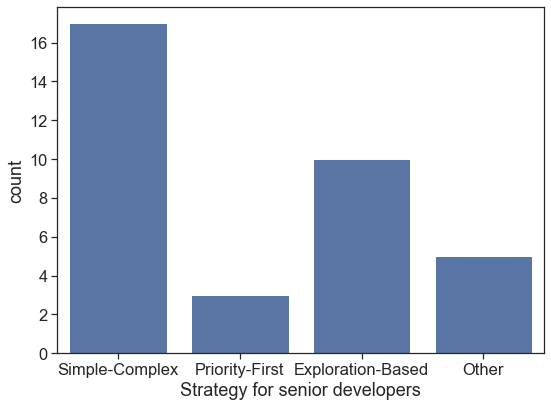} 
       \caption{Manager-reported onboarding strategies for senior developers ($N=35$).}
       \label{fig:manager_strategies_senior}
    \end{subfigure}
    \caption{Survey results about onboarding strategies.}
    \label{fig:survey_strategies}
\end{figure*}
\subsection{About experiences and recommendations}
Fig.~\ref{fig:survey_results} shows that findings in Table~\ref{tab:findings_recommendations} are representative of onboarding experiences of developers and managers from \company{}.
Most developers agree with our interview findings and most managers find our recommendations useful.
In Fig.~\ref{fig:developer}, the average percent of developers who agree or strongly agree with our statements about developer experiences is 88.9\% and the min is 78.5\%.
In Fig.~\ref{fig:manager}, out of managers who are not using the recommended practice at the time,
the average percent of managers who believe that recommended practices are useful for onboarding is 81.9\% and the min is 59.1\%.
Besides, Fig.~\ref{fig:manager_in_practice} shows that on average, 36.8\% managers are already using recommended practices, indicating that 
1) the recommended practices are useful, and 2) recommended practices are actionable.
Therefore, our findings echo with a majority of developers and managers from \company{}.
As \company{} has many divisions with difference engineering culture, these results suggest that our findings can generalize to other software engineering teams and organizations as well.

\subsection{About the three strategies}
Fig.~\ref{fig:manager} shows that the three strategies presented in Section~\ref{sec:strategies} can represent most onboarding strategies that managers use in \company{}.
In Fig.~\ref{fig:developer_strategies}, the three strategies cover 96.3\% onboarding experiences.
Fig.~\ref{fig:manager_strategies_junior} and Fig.~\ref{fig:manager_strategies_senior} show that 94.4\% and 85.7\% managers use one of the three strategies to describe their onboarding strategy. 
Therefore, the three strategies can explain the design intention of most onboarding experiences.

Comparing Fig.~\ref{fig:manager_strategies_junior} and Fig.~\ref{fig:manager_strategies_senior}: 
1)  86.1\% managers onboard junior developers with Simple-Complex strategy as compared to 48.6\% when onboarding senior developers;
2)  27.6\% managers use Exploration-Based strategy when onboarding senior developers as compared to 5.5\% for junior developers.
These two arguments are consistent with our interview findings. 
The comparison also suggests that managers have more diverse onboarding strategies for senior developers. 

\section{Discussion and Related Work}
\label{sec:related}

Onboarding is more broadly studied in other areas, especially management and human resources~\cite{bauer2010onboarding,jacobs2003structured,meyer2017impact,saks2007socialization}.
Nonetheless, domain-specific observations are necessary because software development tasks have unique properties.
For example, open-ended tasks, used in Exploration-Based strategies, are not common in other teams such as service teams.
Besides, our study focuses on onboarding collocated teams, while assigning tasks in distributed teams requires more considerations~\cite{simao2018task}.
Future studies may generalize our findings to distributed teams.

Our work is inspired by Deganais et al.'s study on developers moving into a new landscape~\cite{dagenais2010moving}.
In comparison, some of our observations about learning are consistent with theirs.
We notice, beyond Deganais et al.'s study, that learning the landscape is not the only objective for onboarding,
so we include other indicators such as social connections~\cite{rollag2005getting,ellis2015navigating}.
For example, Deganais et al. recommend early experimentation to facilitate learning;
our study shows that early experimentation is one strategy (Simple-Complex) to onboard new members,
and other strategies (Priority-First and Exploration-Based) could also be used in certain cases.
Similarly, our work echos with Begel \& Simon's research on novice developers~\cite{begel2008novice,begel2008struggles},
while we present a more comprehensive view of longitudinal onboarding around tasks.
In 1998, Sim \& Holt interviewed four developers and summarized seven patterns in four major categories~\cite{sim1998ramp}.
Compared with this work, our study is more complete and modern.
For example, Sim \& Holt's pattern 4 is ``initial tasks were open-ended problems or simple bug repairs''~\cite{sim1998ramp};
we summarize this pattern into two strategies: Exploration-Based and Simple-Complex,
and we have additionally uncovered the Priority-First strategy, which has a close relationship with Agile development method~\cite{beck2001manifesto} that became mainstream after Sim \& Holt's work. 


Our study is closely related to studies on selecting tasks to enter open-source communities~\cite{mendez2018open,steinmacher2015social,steinmacher2015understanding,steinmacher2015systematic,casalnuovo2015developer}.
Our work distinguishes from theirs in two significant ways.
First, we focus on industry, which have major distinctions from open-source communities in terms of motivations, roles, and responsibilities~\cite{ye2003toward}.
Second, we examine a process towards an independent contributor while they focus on entering a community.
As our study shows, onboarding has multiple stages with different objectives, and entering a community/team is just one stage.
Only when all stages are considered together can we explain some strategies, such as Priority-First, that practitioners use to onboard new members 

There are a few other studies that are related to tools to choose tasks for onboarding~\cite{wang2011bug,stanik2018simple}.
One such example is Wang \& Sarma's work on finding the right bug for new members~\cite{wang2011bug}.
Our study provides more guidance for researchers to design and test such automated systems.
For example, Egelman et al.'s recent work on detecting negative interpersonnal interactions from code reviews~\cite{egelman2020pushback}
could be used to support onboarding, as our study suggests that negative feedback in code reviews could hurt a new developer's confidence and leads
to a less successful onboarding experience.


\section{Threats to Validity}
\label{sec:threats}
Our study is a mixed-method case study with interviews and surveys.
We address three types of threats to validity~\cite{wohlin2012experimentation}. 

\paragraph{Construct Validity}
In Section \ref{sec:survey}, we use the percentage of agreements in our survey as an indirect measurement of validity.
We assume that developers and managers who have reported agreement to our results have had experiences or perspectives that are consistent with our results.
In response to concerns about the validity of this measurement, we design surveys to be clear, concrete, and neutral.
For example, we refer to practice recommendations as practices in manager survey to avoid biases;
we also frame questions so that managers are not forced to agree with the usefulness of presented practices.
In both surveys, each question offers participants a place to explain their choices.
From these answers, we have seen no indicator that any question is vague or misleading.

\paragraph{Internal Validity}
In interviews, we ask subjects to recall their recent activities and feelings, which are subject to biases.
To address this, we choose subjects who have joined a new team recently or who have a new member joined recently. 
As a future direction, a longitudinal study that observes developers who are onboarding a team might provide more insights. 
Another threat to validity is that our summary of interviews is subject to researchers' personal biases.
Therefore, most interviews are conducted by two researchers, and interview findings are consolidated among researchers.
Besides, all interviews are recorded so that researchers can refer to the original data when needed.
We have also shared our results with several domain experts within \company{} in the form of write-ups and presentations to avoid biases. 

\paragraph{External Validity}
This study is conducted in one company, which poses a threat to the study's external validity.
While we recognize that similar studies from more companies are desired,
we believe our study should generalize to most modern software engineering teams.
\company{} is a large company with teams working on multiple technical areas serving various customers.
Its teams are self-organized and thus have the autonomy to practice various development methods and to different degrees.
Furthermore, we sampled participants from two company divisions that have different products, businesses, and cultures.     
This sampling method also improves the external validity of our results.


\section{Conclusion and Future directions}
\label{sec:conclusion}

To our best knowledge, this study is the first research work on software developer onboarding tasks and strategies.
By interviewing developers and managers, we present how tasks influence three representative themes of onboarding: learning, confidence, and social interactions.
Our findings lead us to some recommendations for practitioners to improve onboarding processes. 
Furthermore, we have summarized three strategies that managers use to onboard new members.

Our work benefit both researchers and practitioners.
For researchers, we provide observations and insights for software developer onboarding.
For practitioners, we provide a list of recommendations in Table \ref{tab:findings_recommendations}. 
Practitioners from \company{} have validated that these recommendations are useful and practical for onboarding new members.

We encourage future studies to provide quantitative perspectives on onboarding.  
For example, experiments that compare the three onboarding strategies would bring many benefits to the software engineering community.
Our findings also open up venues for empirical and experimental research in onboarding area.
For example, in Section~\ref{sec:strategies}, we hypothesize that \emph{exploration-based onboarding strategies could cause a bad onboarding experience for less experienced developers.}
Such a hypothesis could be validated more rigorously by experiments.
As a case study, our measurements of tasks are subjective based on our interviews, and we encourage future studies to design objective measurements to further assist onboarding.

Onboarding is a fundamental issue correlated with productivity, job satisfaction, retention, creativity, and other desired outcomes of an organization.
With tasks, new members are shaped by the team but also reshape the team.
We envision a theory where we can understand this dynamic, profound, and critical process.
Such a theory would help the community to keep teams, the minimal units of the software engineering industry, efficient, safe, and lively.

\bibliographystyle{IEEEtran}
\bibliography{IEEEabrv,main}
\end{document}